# Dose and Fluence Distributions of the Primary and Secondary Particles in Biological Material Irradiated by $^{252}$Cf Fission Neutrons and d-Be Generated Neutrons

Sungmin Pak and Francis A. Cucinotta[1]

*Department of Health Physics and Diagnostic Sciences, School of Integrated Health Sciences,*

*University of Nevada, Las Vegas, Nevada*

[1] Address for correspondence: University of Nevada, Las Vegas, Integrated Health Sciences, BHS 342, 4505 S. Maryland Pkwy., Las Vegas, Nevada 89154; email: francis.cucinotta@unlv.edu.

# Dose and Fluence Distributions of the Primary and Secondary Particles in Biological Material Irradiated by $^{252}$Cf Fission Neutrons and d-Be Generated Neutrons

Sungmin Pak and Francis A. Cucinotta

## ABSTRACT

For understanding the biological effects of neutrons, predictions of the secondary charged particle distributions by neutron irradiation are needed in biophysical models. We have performed detailed Monte-Carlo simulations using the PHITS computer code of the the dose and fluence spectra of charged particles in the biological materials irradiated by neutron beams with energies below 10 MeV. We compare the results for two different neutron spectra used in radiobiology experiments; the spontaneous fission neutron spectrum of $^{252}$Cf, and a 4 MeV d-Be generated neutron spectrum. The results show that over 90% of the dose and fluence are from secondary protons, which are low energy (<2 MeV) and high LET, and indicate higher secondary charged particle fluence near the surface compared to the deep tissue regions in a mouse. It is also suggested that the different neutron sources considered result in largely similar types of secondary particles with modestly varying fluence distributions.

1. INTRODUCTION

Neutrons are utilized in several academic, industrial, and medical areas, such as neutron activation analysis, nuclear reactors, and Boron Neutron Capture Therapy (BNCT). In proton therapy for cancer treatment, neutrons are a prominent secondary radiation which vary with passive and active methods for proton generation [1], and neutrons and therapy energy protons will produce a variety of secondary charged particle species [2]. Secondary neutrons are present in spaceflight [3, 4], generated mainly by cosmic ray protons, heavy ions, and secondaries including neutrons, incident on spacecraft, atmospheres or tissue structures, which contributes to health risks in space travel [5-7]. It is well known that neutrons produce fast recoil protons and heavy-ion fragments when interacting with matter. Therefore, the secondary particle fluence distribution should be known to comprehend the biological effects of neutrons, and are also needed for modeling studies of neutrons or for proton and heavy ion irradiation with significant secondary neutron generation.

Interest in neutron radiobiology in the 1970's and 1980's had a primary focus on understanding cancer and other health risks of the Atomic-bomb survivors in Japan and for nuclear reactors workers were neutrons present a minor dose contribution compared to gamma-rays, however have a much larger biological effectiveness [8, 9]. In recent years, studies with neutron sources with energies <10 MeV have been reported for mouse tumor induction or cognitive effects [10-12], while the recent tumor studies can be compared to older studies for a variety of mouse strains and tumor types [13-15]. Several neutron sources have been used, ranging from radioisotopes to particle accelerators. One of the most commonly used radioactive isotopes to produce neutrons is $^{252}$Cf, a powerful neutron emitter. Its spontaneous fission neutron spectrum has been studied and measured, which is empirically suggested as $N(E) \propto \exp(-0.88E) \sinh(2.00E)^{1/2}$, where $E$ is the neutron energy in the unit of MeV, by Smith AB, Fields PR, Roberts JH, 1957 [16]. For studying of biological aspects of the neutron beam, the neutron exposure accelerator system for biological effect experiments (NASBEE) is being operated at the National Institute of Radiological Sciences (NIRS) in Japan [17]. The neutron spectrum from 4 MeV deuteron bombardment on beryllium target (d-Be reaction) at this facility is described by Takada M, et al., 2011 [18]. The neutron spectra from 0 to 10 MeV for $^{252}$Cf spontaneous fission neutrons and 4 MeV d-Be generated neutrons are depicted in **Fig. 1**. In this work, we compare the absorbed dose, and neutron, photon, and secondary charged particle fluences during the $^{252}$Cf and d-Be neutrons bombardments on biological material represented by water or muscle tissue.

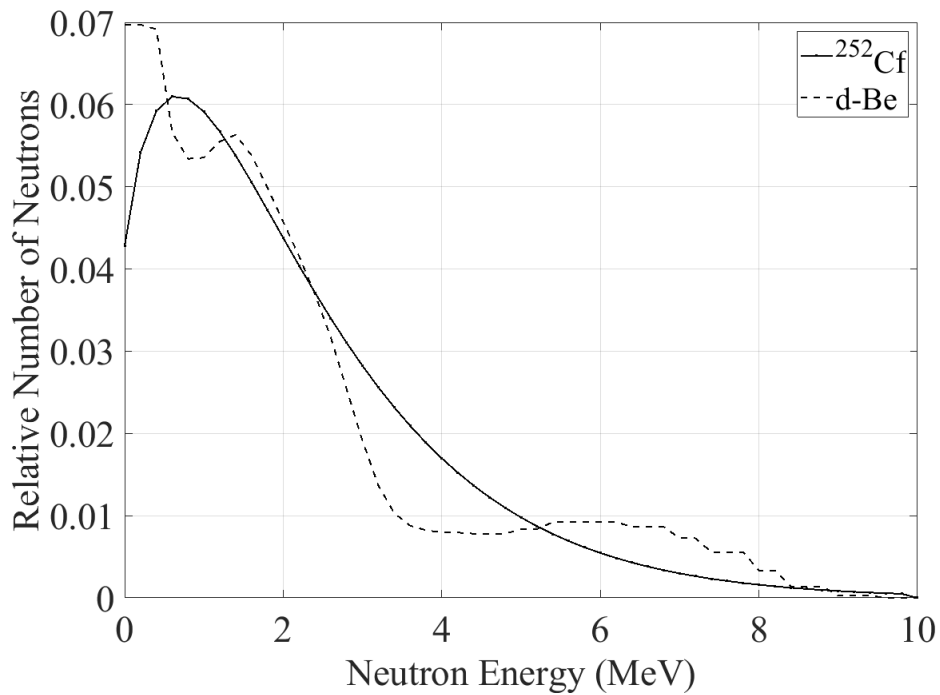

**FIG. 1.** The energy spectra of $^{252}$Cf spontaneous fission neutrons and 4 MeV d-Be generated neutrons. The number of neutrons is normalized to the total number of neutrons from each source.

2. **METHODS**

The Particle and Heavy Ion Transport code System (PHITS) simulates particle movement and interactions between particles and materials using the Monte Carlo method [19, 20]. Among various physics models in PHITS, the Liége intranuclear cascade (INCL) 4.6 model [21] and the Jaeri Quantum Molecular Dynamics (JQMD) model [22-24] for light charged particles and heavy charged particles have been selected, respectively, as they show a good performance for charged particle study [25]. For transportation of charged particles in a low energy range, the ATIMA algorithm has been used [20]. The event generator mode has been utilized for generation of secondary charged particles from low energy neutron reactions [26-28], and the Electron-Gamma Shower 5 (EGS5) algorithm has been chosen for the photon transport description [29]. The latest PHITS (version 3.24) is adopted for this work.

In order to investigate the dose and particle fluences in a mouse, a spherical phantom made-up of water or alternatively tissue material with a mass of 20 grams and a radius of 1.684 cm is chosen as the main target. Two smaller hypothetical spheres with a mass of 116.64 milligrams and a radius of 0.303 cm set as sensitive volumes are located at the center and at the entrance of the mouse to estimate the dose and particle fluences at shallow and deep regions. Here, the volume at the center demonstrates a lung, and the volume at the entrance stands for the region below the surface where the beam is incident on the mouse. Mono-directional neutron beams with a diameter of 4 cm, which are large enough to irradiate the whole body of the mouse, have been generated at a 5 cm distance from the center of the mouse and headed to the target. For the simulation we assume there are no shielding or dense materials in the beamline, and the target is surrounded by air. However, in practical scenarios some shielding approach would be used to reduce photon doses to a neglible amount, and cages to hold mice may be used which could moderate the neutron spectrum.

One hundred million neutrons, i.e., 100,000,000 n/4π(2 cm)² ≈ 1,989,437 n cm⁻² for the initial neutron fluence, are sampled for each neutron spectrum: $^{252}$Cf spontaneous fission and d-Be reaction. Fluences of neutron, photon, and secondary charged particles, and the absorbed doses from each secondary ions at the entire body ("Body"), at the entrance of the body ("Entr"), and at the center of the body ("Lung") have been recorded in the PHITS simulations in water. In addition, the simulation for a tissue target with $^{252}$Cf spontaneous fission neutron spectrum has been conducted to compare the difference between water and tissue. Tissue composition by weight was 10% of H, 65% of O, 18% of C, and 3% of N. The geometry of the PHITS simulation is depicted in **Fig. 2**.

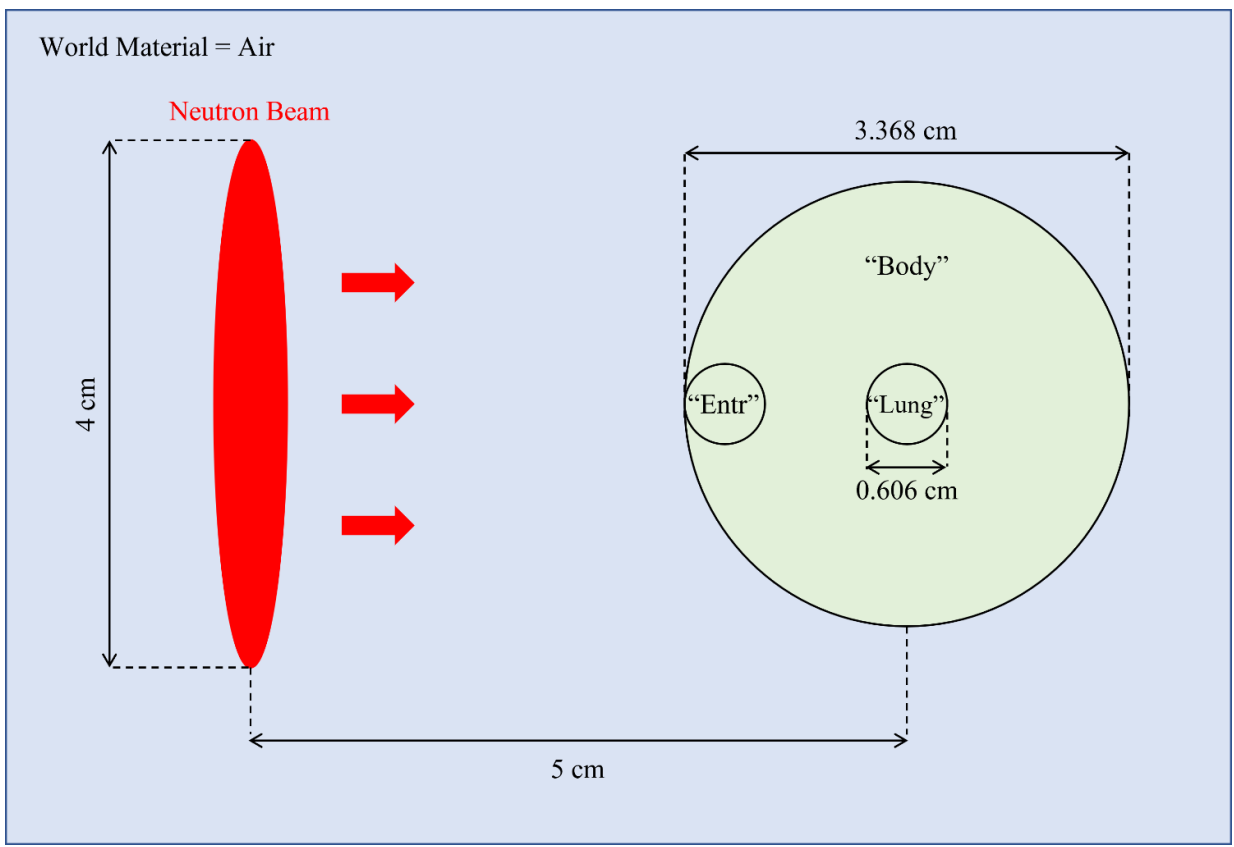

**FIG. 2.** The geometry of the PHITS simulation. Particle doses and fluences at the entire body of the mouse ("Body"), at the entrance of the body ("Entr"), and at the center of the body ("Lung") have been recorded.

## 3. RESULTS

The absorbed doses from one hundred million neutrons (corresponding to 1.989 x 10$^6$ n/cm$^2$ initial neutron fluence) at the entire body of the mouse ("Body"), at the entrance of the mouse ("Entr"), and at the center of the mouse ("Lung") for $^{252}$Cf spontaneous fission neutron spectrum and 4 MeV d-Be generated neutron spectrum are described in **Fig. 3**. The dose at the entrance is highest, and $^1$H ions, $^{16}$O ions, $^4$He ions, and $^{13}$C ions makeup ~99% of the absorbed dose at each volume. The total absorbed dose from all charged particles and the detailed dose contributions of the four major secondary ions are listed in **Table 1**. It has been revealed that, at the same volume, $^{16}$O ions have slightly higher contributions for $^{252}$Cf fission neutrons, while the portions of $^1$H ions, $^4$He ions, and $^{13}$C ions are higher for d-Be generated neutrons. It has also been found that, for the same volume, $^{252}$Cf source delivers a higher dose compared to the d-Be reaction, for the same number of emitted neutrons.

Particle fluence spectra at the entire body, at the entrance, and at the center for both neutron sources are given in **Figures 4 to 6**. With one hundred million initial neutrons, the statistical error is revealed to be ignorable (<<1%) for the four major particles: $^1$H ions, $^{16}$O ions, $^4$He ions, and $^{13}$C ions, while higher statistical errors occur for other particles. The fluences are normalized to one cGy of surface neutron dose, i.e., the dose at the entrance of the mouse ("Entr") for each neutron source. Of note is that the absorbed doses at each volume are not the same, with their predicted values found in **Table 1**.

The major secondary particles are $^1$H ions (protons) and $^{16}$O ions, which are the components of the target material produced in elastic scattering. The $^1$H, $^4$He and other ions are produced from inelastic scattering or spallation reactions during the interaction between neutrons and the water or tissue molecules. Neutron capture on $^{12}$C or $^{16}$O leads to a small $^{13}$C or $^{17}$O flux, respectively. The fluences of $^2$H ions (deuterons) and $^{17}$O ions exist at all three sensitive volumes ("Body", "Entr", and "Lung") but are not significant. Negligible amounts of $^3$H (tritium), $^{11}$B, $^{12}$C, $^{14}$C, $^{14}$N, $^{15}$N ions have been recorded only for the larger volume ("Body"). The major secondary charged particle fluences have been revealed to be highest near the surface ("Entr"). The fluence distributions for both neutron sources are revealed to be similar but slightly different. For example, the fluence of $^{16}$O ions is higher for $^{252}$Cf neutrons, while the d-Be reaction has higher charged particle fluences for $^1$H, $^4$He, and $^{13}$C ions.

It is shown that most neutrons penetrate the target without inelastic collisions, which results in remarkably high neutron fluence compared to any other particles. Unlike the secondary charged ions, the fluences of neutrons and photons tend to be higher at the center of the body ("Lung") compared to other sensitive volumes. The higher neutron fluence for d-Be generated neutrons in the low energy range is considered due to the prominent low energy (< 1 MeV) initial neutron fluence for this source.

Comparisons between water and tissue targets for $^{252}$Cf neutron source are shown in **Fig. 7, Fig. 8, and in Supplementary Figures, Fig.'s S1 and S2.** The absorbed dose in tissue is revealed to be ~ 4.5% smaller compared to water for the identical initial neutron spectra. As tissue contains carbon and nitrogen components, $^{12}$C ions and $^{14}$N ions are considered amongst the major secondary particles in tissue exposed to neutrons, however are negligible in water. The fluences of the minor secondary ions, including $^3$H, $^9$Be, $^{10}$Be, $^{11}$B, $^{14}$C, $^{14}$N, and $^{15}$N ions, are found to be higher in tissue than water; however, they do not make significant contributions to the absorbed dose. Particle fluence spectra of $^1$H, $^{16}$O, $^4$He, and $^{13}$C ions are similar in both materials. The computation time for both materials is identical, which using 35 processors of Intel® Xeon® CPU E5-2697 v2 @ 2.70 Ghz and 128 GB RAM requires ~ 7 days.

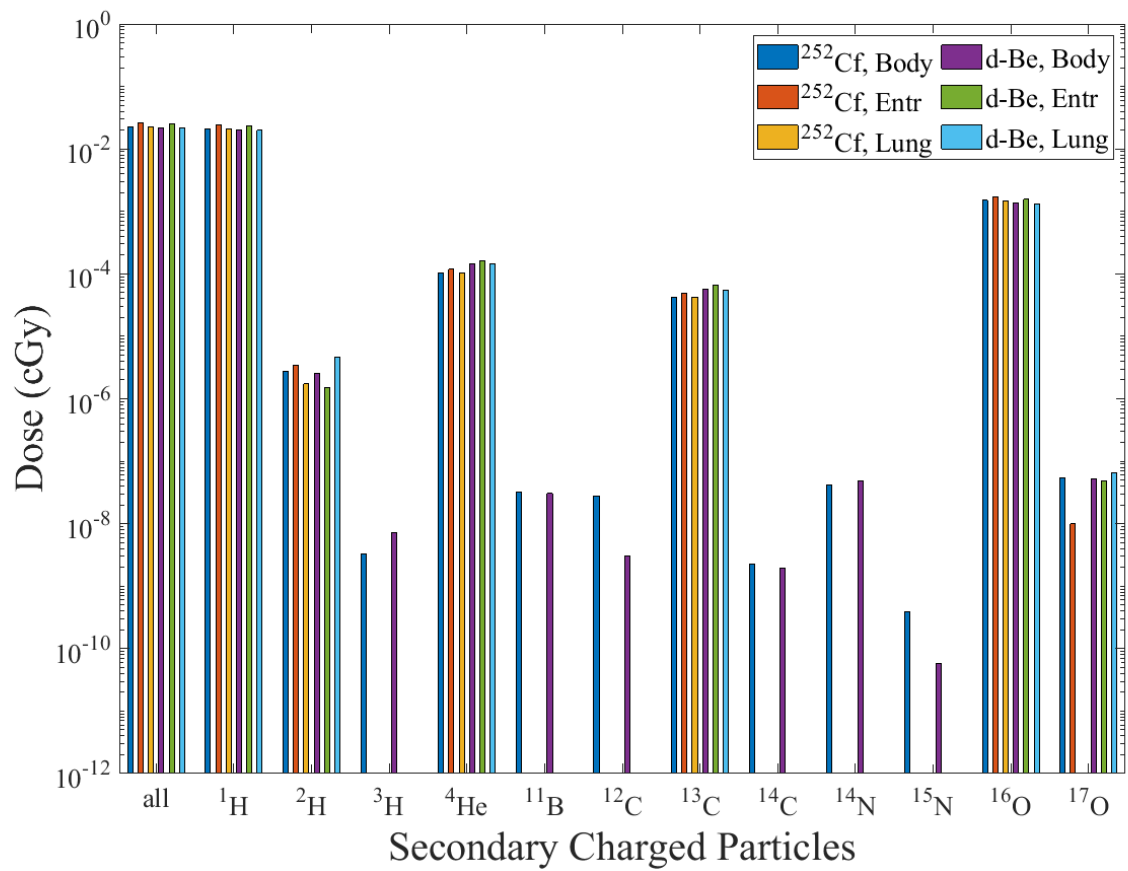

**FIG. 3.** Absorbed dose at the whole body ("Body"), at the entrance of the body ("Entr"), and at the center of the body ("Lung") from the secondary ions for exposure to $^{252}$Cf spontaneous fission neutrons ("$^{252}$Cf") and 4 MeV d-Be generated neutrons ("d-Be").

**Table 1.** Total absorbed doses from one hundred million neutrons (1.989x10$^6$ n/cm$^2$) and dose contributions of the major secondary charged particles at the whole body ("Body"), at the entrance of the body ("Entr"), and at the center of the body ("Lung") for exposure to $^{252}$Cf spontaneous fission neutrons ("$^{252}$Cf") and 4 MeV d-Be generated neutrons ("d-Be").

|  | Total Absorbed Dose (cGy) | Dose Contributions (%) | | | |
| --- | --- | --- | --- | --- | --- |
|  |  | $^1$H | $^{16}$O | $^4$He | $^{13}$C |
| $^{252}$Cf, Body | 0.022565 | 92.701 | 6.630 | 0.462 | 0.186 |
| $^{252}$Cf, Entr | 0.026029 | 92.727 | 6.608 | 0.457 | 0.191 |
| $^{252}$Cf, Lung | 0.022251 | 92.832 | 6.484 | 0.472 | 0.190 |
| d-Be, Body | 0.021733 | 92.804 | 6.254 | 0.656 | 0.260 |
| d-Be, Entr | 0.025113 | 92.868 | 6.206 | 0.645 | 0.263 |
| d-Be, Lung | 0.021390 | 92.880 | 6.146 | 0.670 | 0.258 |

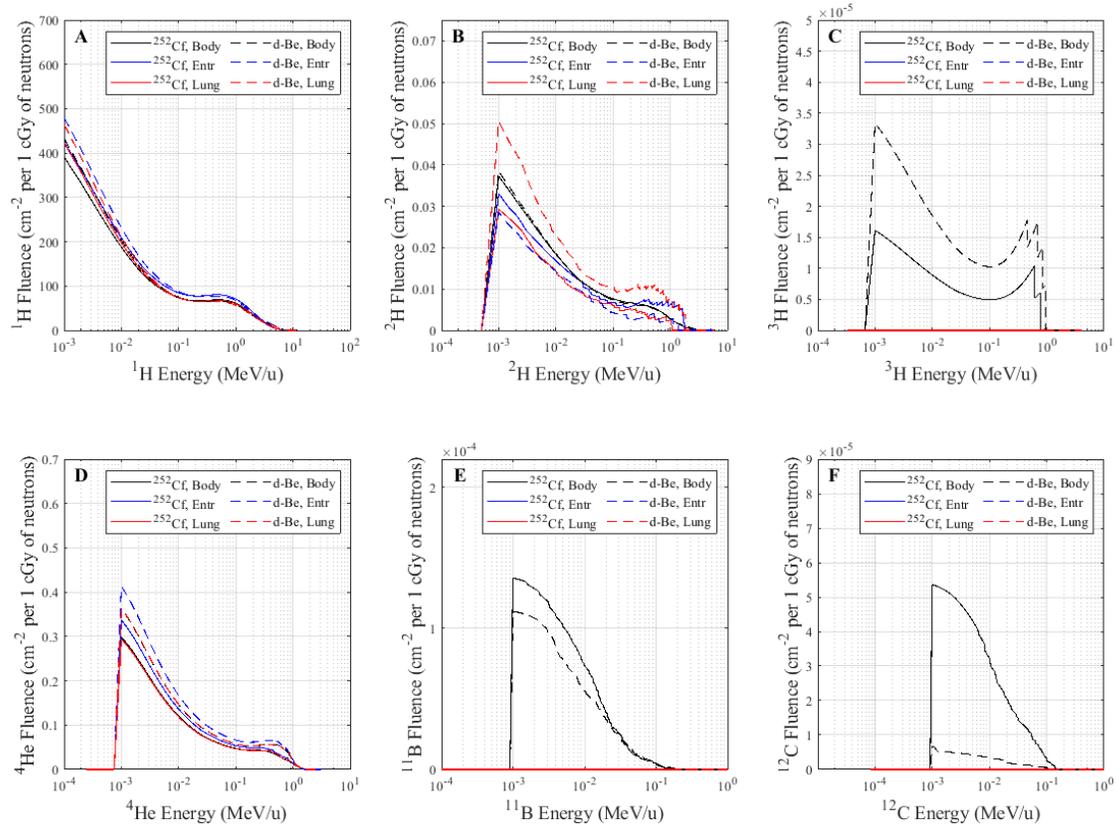

**FIG. 4.** Fluence of several species of charged particles at the whole body ("Body"), at the entrance of the body ("Entr"), and at the center of the body ("Lung") for exposure to $^{252}$Cf spontaneous fission neutrons ("$^{252}$Cf") and 4 MeV d-Be generated neutrons ("d-Be"). The particle fluence is normalized by 1 cGy of dose at the entrance. Panels A), B), C), D), E) and F) are spectrum of $^1$H, $^2$H, $^3$H, $^4$He, $^{11}$B, and $^{12}$C ions, respectively.

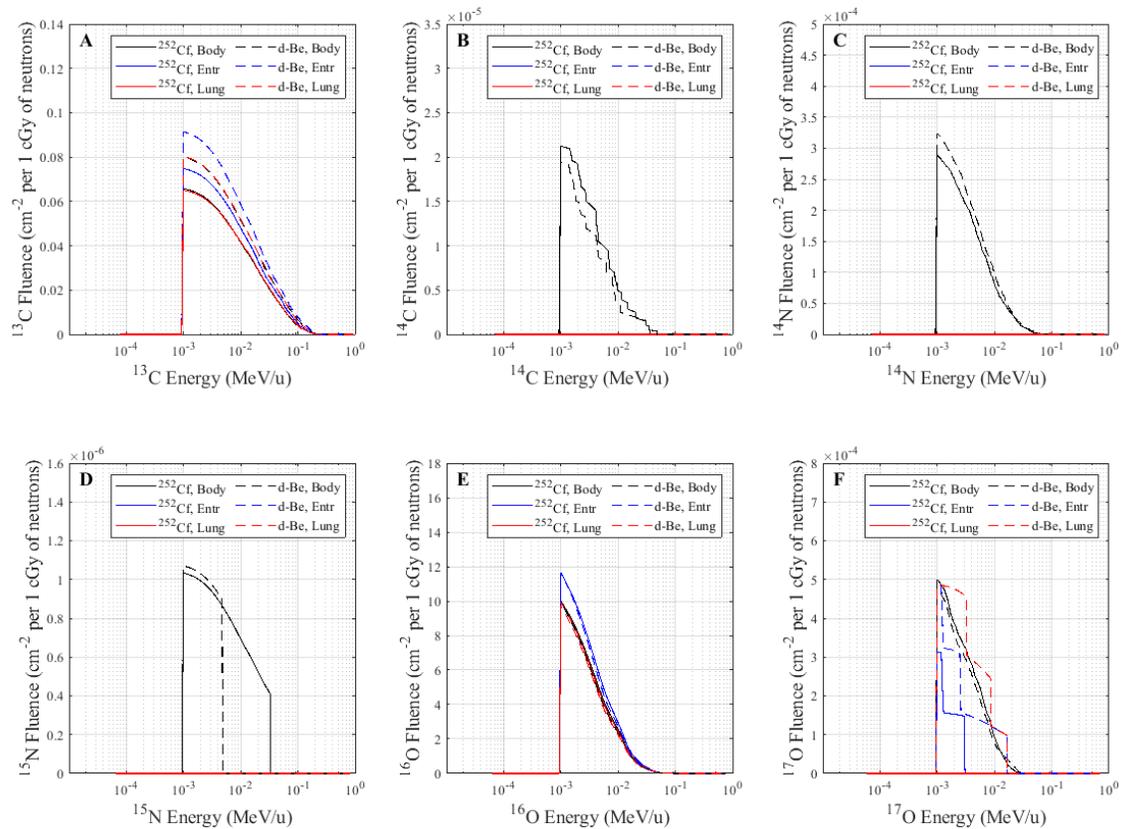

**FIG. 5.** Fluence of several species of charged particles at the whole body ("Body"), at the entrance of the body ("Entr"), and at the center of the body ("Lung") for exposure to $^{252}$Cf spontaneous fission neutrons ("$^{252}$Cf") and 4 MeV d-Be generated neutrons ("d-Be"). The particle fluence is normalized by 1 cGy of dose at the entrance. Panels A), B), C), D), E) and F) are spectrum of $^{13}$C, $^{14}$C, $^{14}$N, $^{15}$N, $^{16}$O and $^{17}$O ions, respectively.

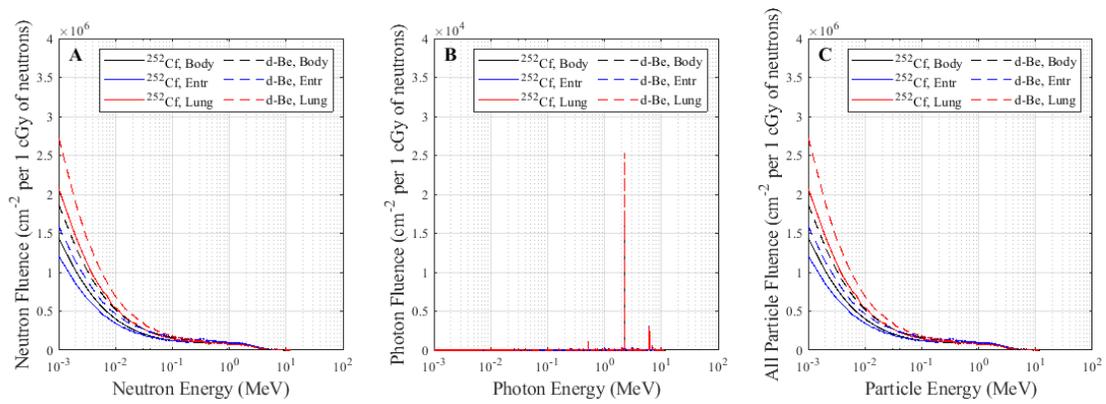

**FIG. 6.** Fluence of neutrons, photons, and all particles at the whole body ("Body"), at the entrance of the body ("Entr"), and at the center of the body ("Lung") for exposure to $^{252}$Cf spontaneous fission neutrons ("$^{252}$Cf") and 4 MeV d-Be generated neutrons ("d-Be"). All particle fluence is composed of all charged and non-charged particle fluences. The particle fluence is normalized by 1 cGy of dose at the entrance. Panels A), B), and C) are spectrum of neutrons, photons, and all particles, respectively.

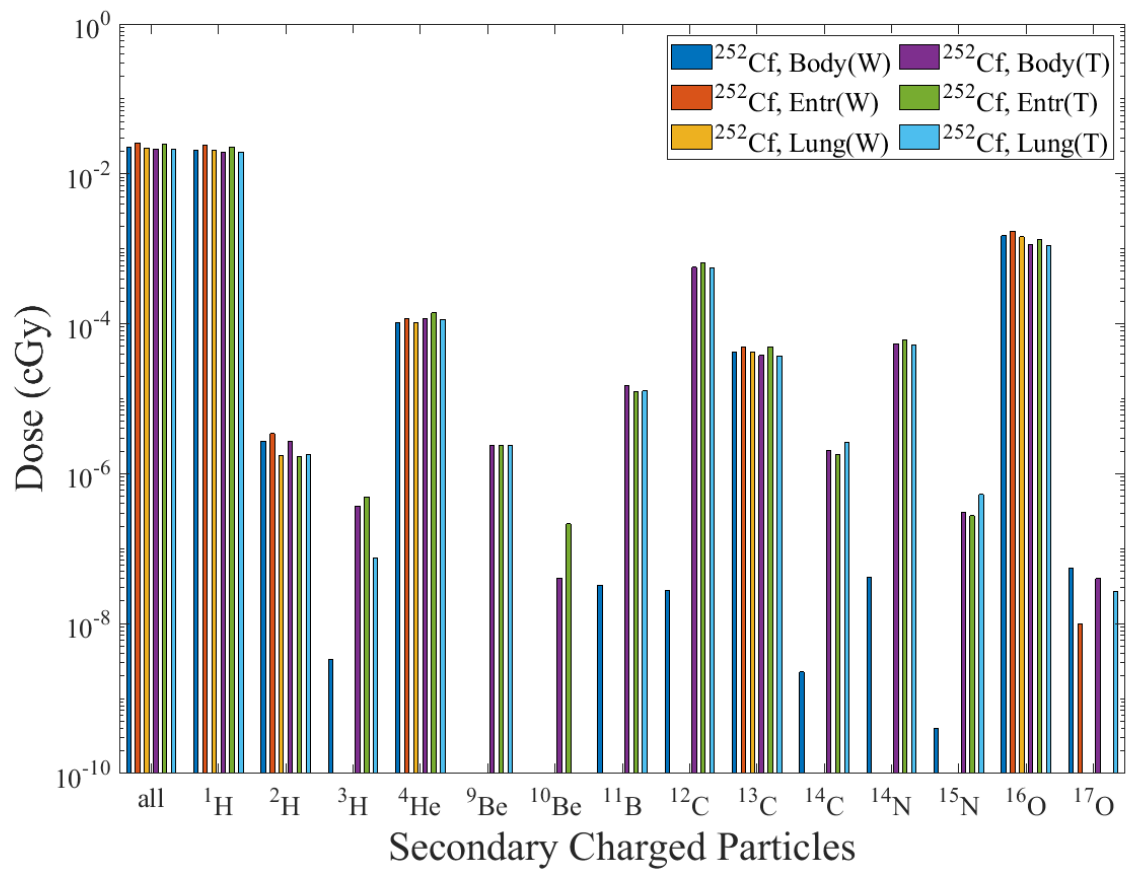

**FIG. 7.** Absorbed dose at the whole body ("Body"), at the entrance of the body ("Entr"), and at the center of the body ("Lung") of water ("W") and tissue ("T") from the secondary ions for $^{252}$Cf spontaneous fission neutron ("$^{252}$Cf") irradiation.

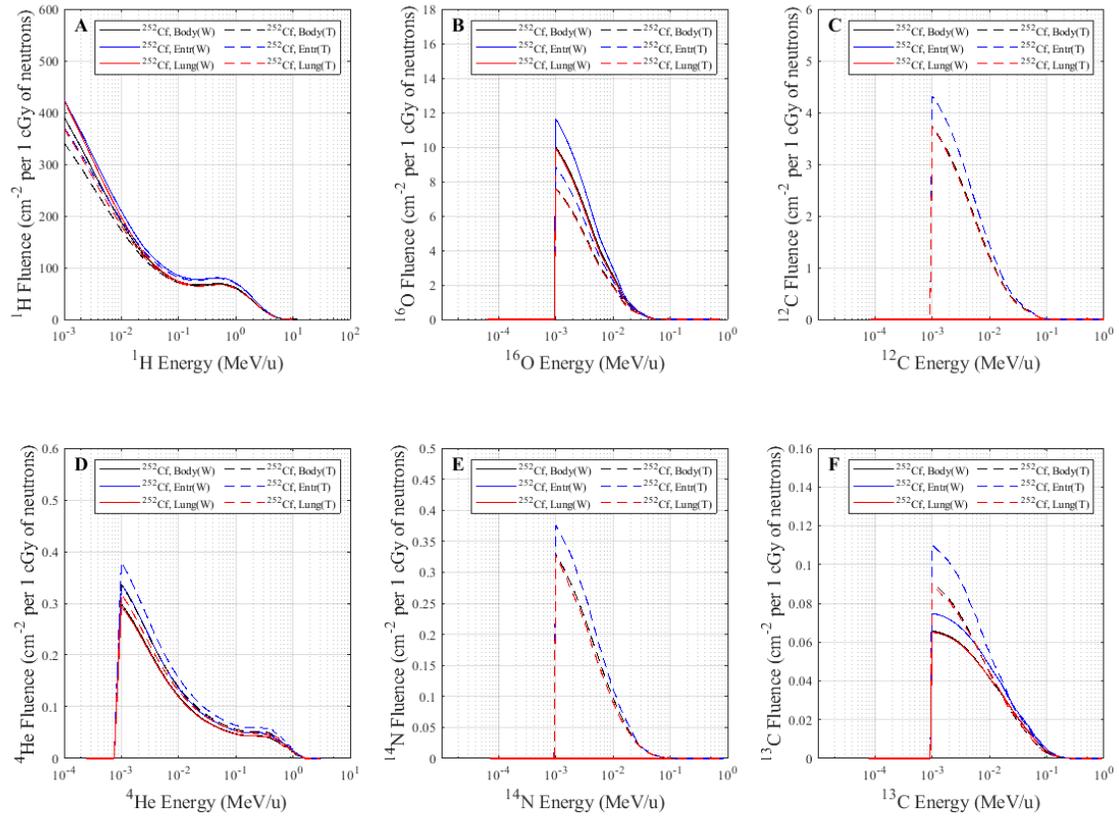

**FIG. 8.** Fluence of several species of charged particles the whole body ("Body"), at the entrance of the body ("Entr"), and at the center of the body ("Lung") of water ("W") and tissue ("T") for $^{252}$Cf spontaneous fission neutron ("$^{252}$Cf") irradiation. The selected particles are considered the major secondary particles in tissue. The particle fluence is normalized by 1 cGy of dose at the entrance. Panels A), B), C), D), E) and F) are $^1$H, $^{16}$O, $^{12}$C, $^4$He, $^{14}$N, and $^{13}$C ions, respectively.

## 4. DISCUSSION

The use of neutron sources with low energies (<10 MeV) has found renewed interest in cancer and CNS effect radiobiology [10-12, 30, 31], and older studies remain as a valuable source understanding high LET tumorigenesis [13-15]. In comparison, use of higher energy neutron sources reduces the average LET values because of the production of higher energy charged particles which require more complex transport model descriptions due to the larger number of spallation reactions that would occur as energy is increased. The secondary particle doses and fluences in the biological material have been compared between $^{252}$Cf spontaneous fission neutron irradiation and d-Be reaction generated neutron irradiation. The fluence distributions of the major four secondary ions ($^1$H, $^{16}$O, $^4$He, and $^{13}$C) are similar but slightly different for these two sources. It has been found that there are larger particle fluences and doses of secondary ions at shallow tissues compared to deep regions. Both neutron sources are suggested to generate the same type of minor secondary particles, however are ignorable due to their small fluence. $^2$H ions and $^{17}$O ions have higher contributions than other minor secondary particles; however, further simulations with more neutrons are needed to understand their fluences and dose contributions. It has been found that the same initial neutron fluence from the $^{252}$Cf source gives a higher dose at the biological material because there are more low-energy neutrons in the d-Be neutron spectrum. The major particle fluence distribution in water and tissue for the same neutron source are shown to be similar, while less absorbed dose and more secondary particle contributions other than $^1$H have been measured in tissue.

We did not consider incident photons in our analysis. In practical application photons are likely shielded to a neglible dose contribution in studying neutron effects [32,33]. The PHITS code predicted very small secondary photon dose contributions (<0.03%) for the geometry and tissue volumes considered. In future work the calculations made here could be augmented with photon shielding or to consider cages that hold mice or other equipment in an exposure room. The photon dose contribution will play a larger role for acute effects where the neutron relative biological effectiveness (RBE) factors are modest (<5), however as noted by Storer et al. [13] can likely be ignored for neutron carcinogenesis studies where RBE's are found to be large. If photon doses are ignored in biophysical models it is important that dosimetry used in experiments report both the photon and neutron dose.

The energy spectrum of heavy ions is extremely low (<0.1 MeV/u) as shown in our predictions. This presents the question of what role do they play in neutron biological effects. First, they present a minor contribution to the neutron doses (Table 1). Secondly, the energy deposition from such low energy heavy ions in causing biological damage may be reduced from description based on LET due to their narrow track structures (<20 nm), and short-ranges in tissue. The narrow track structures have been described as leading to thin-down effect for heavy ions [34, 35], which localizes their effects due to the absence of a delta-ray penumbra. The range of $^{16}$O at 0.01 and 0.1 MeV/u in tissue is ~0.01 and 0.4 microns, respectively, and are energies where nuclear stopping contributes to the LET [36]. The above factors suggests that very little effect is caused by a heavy ion that is produced by a neutron unless it is created quite close to an important biological structure such as chromatin. However, such targets make-up a minor volume in tissue considering inter-cellular spacing [37], the cytoplasm and other cell nucleus components. Therefore, their effect can largely be ignored in comparison to damage caused by secondary protons. The extreme localization of such low energy heavy ions suggests short-comings in describing neutron biological effectiveness using LET or microdosimetry approaches, with the latter considering energy deposition in volumes with diameters of ~1 micron, while ignoring the spatial distribution of the energy deposition [33]. In previous work [6, 7] we used a track structure model to estimate neutron effects where we assumed biological effects are dominated by the more energetic high LET protons produced by neutrons using the spectrum estimates from Edwards and Dennis [38]. The results of the present study will be considered to study possible uncertainties due to the particle spectra in track structure model predictions and study possible differences in mouse tumor induction for the neutron sources considered.

In conclusion, our study of two neutron sources in recent use for radiobiology studies detailed the dose and fluence distributions of the secondary ions produced by the sources. The results show that the biological effects from various neutron sources may not be the same, however the differences in proton spectrum are modest which suggest differences are not large. There are different types of neutron exposures, including near nuclear reactors, and secondary neutrons in medical treatments and space missions. Further research is needed to understand the health effects of the diversity of neutron exposure cases, while the calculation presented here should aid in understanding the complexities of the mixed-field produced by neutrons and supports the application of track structure models of biological effects.

**ACKNOWLEDGMENTS**

Funding support was by UNLV and the National Institute of Health-National Cancer Institute (NIH-NCI) Grant 1RO1CA208526-01.

SUPPLEMENTARY INFORMATION

**Fig. S1.** Fluence of several species of charged particles at the whole body ("Body"), at the entrance of the body ("Entr"), and at the center of the body ("Lung") of water ("W") and tissue ("T") for $^{252}$Cf spontaneous fission neutron ("$^{252}$Cf") irradiation. The particle fluence is normalized by 1 cGy of dose at the entrance. Panels A), B), C), D), E) and F) are $^2$H, $^3$H, $^9$Be, $^{10}$Be, $^{11}$B, and $^{14}$C ions, respectively.

**Fig. S2.** Fluence of several species of charged particles, neutrons, photons, all ions, and all particles at the whole body ("Body"), at the entrance of the body ("Entr"), and at the center of the body ("Lung") of water ("W") and tissue ("T") for $^{252}$Cf spontaneous fission neutron ("$^{252}$Cf") irradiation. Ion fluence includes $^1$H and heavier ion fluences, while all particle fluence is composed of all charged and non-charged particle fluences. The particle fluence is normalized by 1 cGy of dose at the entrance. Panels A), B), C), D), E), and F) are spectrum of $^{15}$N, $^{17}$O, all ions, neutrons, photons, and all particles, respectively.

**Supplementary File**

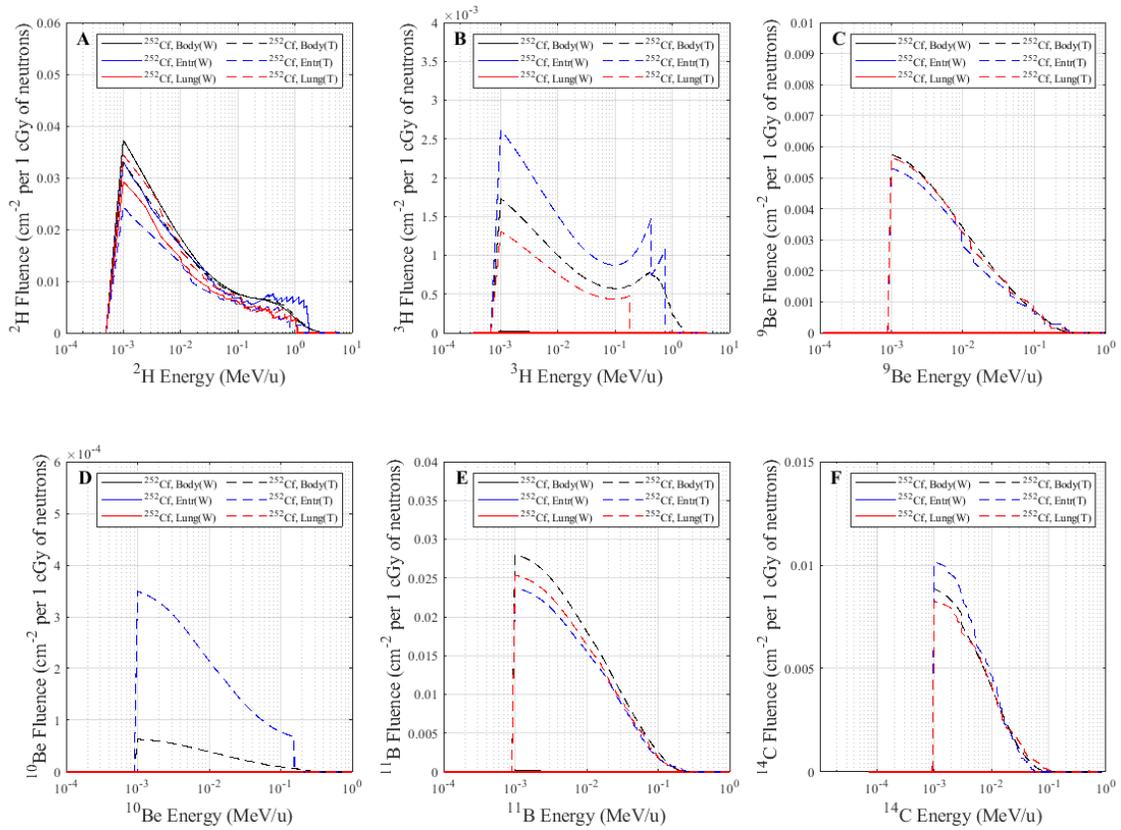

**FIG. S1.** Fluence of several species of charged particles at the whole body ("Body"), at the entrance of the body ("Entr"), and at the center of the body ("Lung") of water ("W") and tissue ("T") for $^{252}$Cf spontaneous fission neutron ("$^{252}$Cf") irradiation. The particle fluence is normalized by 1 cGy of dose at the entrance. Panels A), B), C), D), E) and F) are $^2$H, $^3$H, $^9$Be, $^{10}$Be, $^{11}$B, and $^{14}$C ions, respectively.

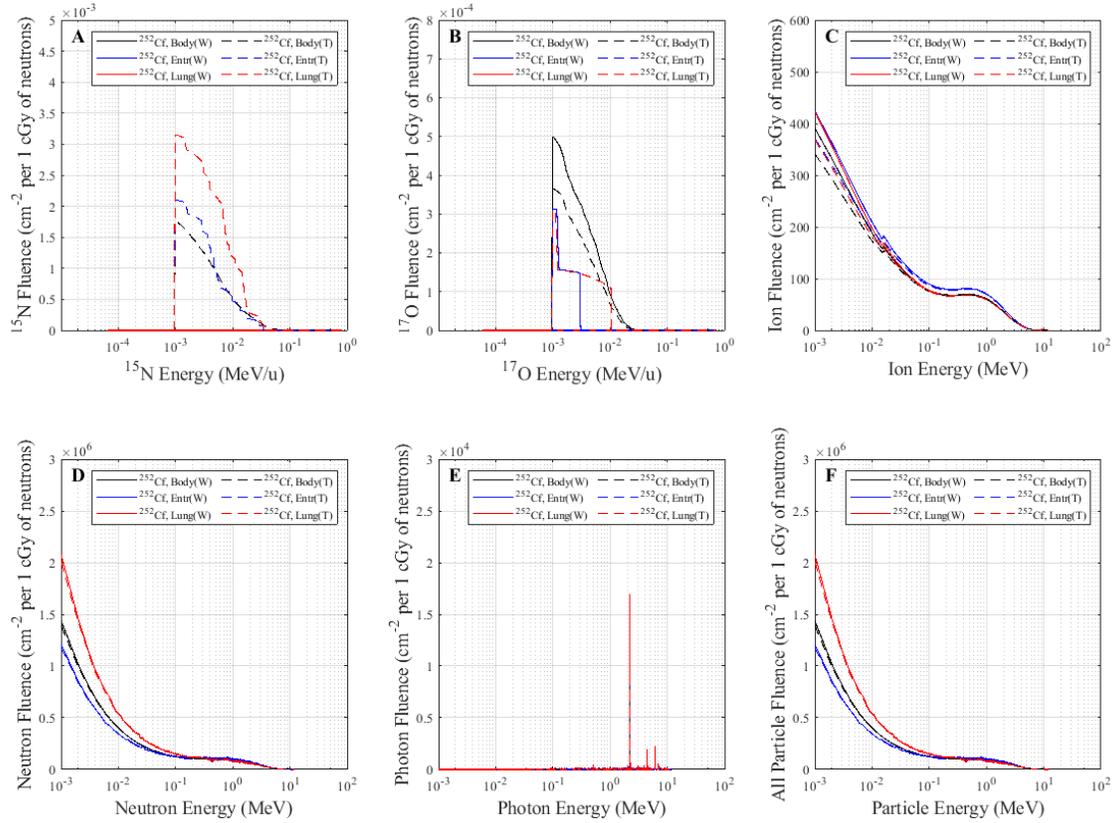

**FIG. S2.** Fluence of several species of charged particles, neutrons, photons, all ions, and all particles at the whole body ("Body"), at the entrance of the body ("Entr"), and at the center of the body ("Lung") of water ("W") and tissue ("T") for $^{252}$Cf spontaneous fission neutron ("$^{252}$Cf") irradiation. Ion fluence includes $^1$H and heavier ion fluences, while all particle fluence is composed of all charged and non-charged particle fluences. The particle fluence is normalized by 1 cGy of dose at the entrance. Panels A), B), C), D), E), and F) are spectrum of $^{15}$N, $^{17}$O, all ions, neutrons, photons, and all particles, respectively.